\documentclass[reprint,
superscriptaddress,
 amsmath,amssymb,
 aps,
prb,
]{revtex4-2}

\usepackage{graphicx}
\usepackage{dcolumn}
\usepackage{bm}

\usepackage[T2A]{fontenc}
\usepackage[utf8]{inputenc}
\usepackage[english]{babel}

\newcommand{\MBT}{MnBi$_2$Te$_4$}
\newcommand{\Pyat}{Mn$_{1-x}$Pb$_x$(Bi$_{1-y}$Sb$_y$)$_2$Te$_4$}

\begin{document}

\preprint{APS/123-QED}

\title{Tuning topological phase and Dirac point position via Pb and Sb substitution in Mn$_{1-x}$Pb$_x$(Bi$_{1-y}$Sb$_y$)$_2$Te$_4$}

\author{T.\,P.~Makarova}
\email{Contact author: i@tmakarova.ru}
\affiliation{Saint Petersburg State University, 198504 Saint Petersburg, Russia}
\author{D.\,A.~Estyunin}
 \affiliation{Saint Petersburg State University, 198504 Saint Petersburg, Russia}
 \affiliation{Center for Advanced Mesoscience and Nanotechnology,
Moscow Institute of Physics and Technology, 141700 Dolgoprudny, Russia}
\author{V.\,A.~Golyashov}
\affiliation{Saint Petersburg State University, 198504 Saint Petersburg, Russia}
\affiliation{Rzhanov Institute of Semiconductor Physics, Siberian Branch, Russian Academy of Sciences, Novosibirsk 630090, Russia}
\affiliation{Synchrotron radiation facility SKIF, Boreskov Institute of Catalysis, Siberian Branch, Russian Academy of Sciences, Kol’tsovo, 630559, Russia}
\author{K.\,A.~Kokh}
\affiliation{Saint Petersburg State University, 198504 Saint Petersburg, Russia}
\affiliation{Sobolev Institute of Geology and Mineralogy, Siberian Branch, Russian Academy of Sciences, Novosibirsk 630090, Russia}
\author{O.\,E.~Tereshchenko}
\affiliation{Saint Petersburg State University, 198504 Saint Petersburg, Russia}
\affiliation{Rzhanov Institute of Semiconductor Physics, Siberian Branch, Russian Academy of Sciences, Novosibirsk 630090, Russia}
\affiliation{Synchrotron radiation facility SKIF, Boreskov Institute of Catalysis, Siberian Branch, Russian Academy of Sciences, Kol’tsovo, 630559, Russia}
\author{A.\,S.~Frolov}
\affiliation{Saint Petersburg State University, 198504 Saint Petersburg, Russia}
\affiliation{Center for Advanced Mesoscience and Nanotechnology,
Moscow Institute of Physics and Technology, 141700 Dolgoprudny, Russia}
\author{S.~Ideta}
\affiliation{Research Institute for Synchrotron Radiation Science (HiSOR), Hiroshima University, Hiroshima 739-0046, Japan}
\author{Y.~Kumar}
\affiliation{Research Institute for Synchrotron Radiation Science (HiSOR), Hiroshima University, Hiroshima 739-0046, Japan}
\affiliation{Graduate School of Advanced Science and Engineering, Hiroshima University, Higashi-Hiroshima 739-8526, Japan}
\author{K.~Shimada}
\affiliation{Research Institute for Synchrotron Radiation Science (HiSOR), Hiroshima University, Hiroshima 739-0046, Japan}
\affiliation{International Institute for Sustainability with Knotted Chiral Meta Matter (WPI-SKCM$^2$), Hiroshima University, Higashi-Hiroshima 739-8526, Japan}
\affiliation{Research Institute for Semiconductor Engineering, Hiroshima University (RISE), Higashi-Hiroshima 739-8527, Japan}
\author{A.\,M.~Shikin}
 \affiliation{Saint Petersburg State University, 198504 Saint Petersburg, Russia}

\date{\today}

\begin{abstract}

This study presents a systematic investigation of \Pyat~crystals over a wide range of concentrations (x = 10–60\%, y = 5–60\%). It was found that the value of the bulk band gap is determined exclusively by the Pb concentration and it closes at Pb~40–50\%, which corresponds to a topological phase transition. The position of the Dirac point is determined by the Pb/Sb ratio, rather than the absolute Sb content. The magnetic properties depend on the dilution of the Mn sublattice by Pb and are weakly sensitive to Sb. We show that the simultaneous substitution of Mn and Bi allows independent control of the topological phase and the position of the Fermi level.

\end{abstract}

\maketitle


\section{\label{sec:level1}Introduction}

\MBT~is one of the most extensively studied magnetic topological systems due to its combination of antiferromagnetic order and non-trivial electronic topology~\cite{otrokov2019prediction}. The quantum anomalous Hall effect (QAHE)~\cite{deng2020quantum}, the axion insulator state~\cite{gao2021layer}, and other unique electronic phases~\cite{li2024dissipationless,liu2020robust} have been predicted in this material. The aforementioned properties of \MBT~render it a promising material for spintronics and low-energy electronic devices~\cite{guo2024quantum,an2021nanodevices,tokura2022quantum}.

The electronic structure of \MBT~is characterized by the presence of topological surface states (TSS) that exhibit a Dirac cone \cite{lee2019spin}. In order to observe QAHE, it is necessary for the Fermi level to lie within the TSS band gap~\cite{li2024recent}. In the case of bulk \MBT~crystals, two factors prevent this condition. 

Firstly, the high concentration of Mn$_{Bi}$ antisite defects (i.e., Mn atoms at the Bi sites) leads to significant n-type doping. This results in the Fermi level lying above the bottom of the conduction band, classifying the material as a degenerate n-type semiconductor with an electron concentration of the order of $\sim10^{20}$~cm$^{-3}$~\cite{yan2019crystal,ji2021detection,liu2020robust,lei2020surface}. This behavior is common to all known magnetic topological insulators (TI) in this family of \MBT-based compounds. Even if doping results in the Fermi level being in the band gap, the high electron concentration persists. This can affect the mobility of charge carriers and other parameters; however, for the observation of the QAHE, the decisive factor is that the Fermi level lies within the gap.

Secondly, the value of the gap in the TSS can vary significantly from sample to sample \cite{shikin2020nature,shikin2021sample}. One of the factors explaining this difference is the presence of Mn$_{Bi}$ antisite defects \cite{garnica2022native}. Therefore, to realize devices based on \MBT, two independent challenges must be solved: a sufficiently wide band gap with the Fermi level inside it, and finding materials where the band gap width is more reproducible and less affected by defects.

The electronic and magnetic properties of \MBT~can be modulated by altering the temperature, the external magnetic field, or by applying hydrostatic pressure. It has been established that pressure enables effective modification of interatomic distances and bulk band gap, while also inducing topological quantum phase transitions (e.g., from TIs to a trivial insulator or a Weyl semimetal, or, in the case of PbBi$_2$Te$_4$ and related compounds, to a superconductor)~\cite{guo2021pressure,xu2022hydrostatic,matsumoto2018data}. Temperature affects the magnetic ordering: below 24~K, \MBT~exhibits A-type antiferromagnetism~\cite{ding2020crystal}.

In contrast to the modulation of temperature and pressure, which can be directly altered during the experimental process, the chemical substitution of atoms during the synthesis is a critical factor that determines the material's properties. The substitution of Bi with Sb in non-magnetic TIs (e.g., Bi$_2$Se$_3$, Bi$_2$Te$_3$) leads to p-type doping and a gradual transition of the system to the trivial phase~\cite{zhang2011band}, but the Dirac cone remains gapless up to the transition due to preserved time-reversal symmetry. In \MBT, substituting Bi with Sb also induces p-type doping and shifts the Fermi level; however, unlike in non-magnetic TIs, it opens a reproducible and tunable gap in the TSS. Notably, undoped crystals show strong sample‑to‑sample variations of the gap, but even at low Sb concentrations the TSS band gap reaches 100~meV and increases monotonically with dopant content~\cite{ma2021realization}. Moreover, this gap persists above the Néel temperature and is insensitive to the magnetic transition.

Furthermore, Sb-doped \MBT~exhibits a hedgehog-like spin texture of the TSS, with oppositely oriented out-of-plane spins at the Dirac gap~\cite{zeng2025hedgehog}. Density functional theory calculations indicate that the system remains topologically nontrivial over the entire studied Sb range. Thus, substituting Bi with Sb in \MBT~provides a convenient way to control the value of the band gap in the TSS and the Fermi level position while preserving the topological properties.

Another approach to modifying the properties of \MBT~is to replace the magnetic Mn atoms with non-magnetic Group IV elements (Ge, Sn, Pb)~\cite{frolov2024magnetic,estyunina2023evolution,shikin2025phase,qian2022magnetic,zhu2021magnetic,tarasov2023topological}. In particular, substituting Mn with Pb leads to the closure of the bulk band gap and a topological phase transition (from TI to a semimetal/trivial insulator and back to TI) at certain Pb concentrations~\cite{estyunin2025electronic}. Moreover, in (Mn,Pb)Bi$_2$Te$_4$, the antiferromagnetic order is preserved, and both the Néel temperature and the spin‑flop transition field decrease monotonically with increasing Pb concentration~\cite{estyunin2023comparative}. At low Pb concentrations ($\sim$10\%), the Néel temperature and the spin‑flop field can even increase compared to pure \MBT. Nevertheless, over the entire studied Pb concentration range, the system remains metallic: the Fermi level lies outside the band gap, which prevents the observation of the QAHE. Substituting Bi with Sb in \MBT~allows effective tuning of the Fermi level; however, it is important to preserve the magnetic and topological properties that arise from Mn/Pb substitution. Therefore, it is necessary to determine whether the combined (Mn/Pb) and (Bi/Sb) substitutions can simultaneously preserve these properties and result in the Fermi level lying inside the band gap. The Pb and Sb concentrations that control the Dirac point position and the Fermi level should also be identified.

The combined substitution of (Mn/Pb) and (Bi/Sb) in \MBT~has not been systematically investigated over a wide range of concentrations. Such combined substitution allows independent tuning of the Fermi level (via Sb) and the magnetic and topological properties (via Pb), which is particularly important for controlling topological phase transitions. Furthermore, this approach is promising for thermoelectric applications because it enables controlled changes in electronic conductivity. In this work, we study the effect of substituting Mn with Pb and Bi with Sb on the magnetic and topological properties of the system over a wide concentration range ($x = 10-60$\%, $y = 5-60$\%). We explore the electronic structure and the Fermi level position, determine the conditions for a topological phase transition, and characterize the magnetic ordering (type, Néel temperature, and spin‑flop transition fields).

\section{\label{sec:level2}Results and Discussion}

To analyse changes in the electronic structure of \Pyat~with varying Pb and Sb concentrations, angle-resolved photoemission spectroscopy (ARPES) measurements were performed. We used photon energies of 21.2~eV to study the bulk band structure and 6.3~eV to investigate the TSS.

\subsection{ARPES at photon energy of $h\nu=21.2$~eV}
 
\begin{figure*}
\includegraphics[width=0.7\textwidth]{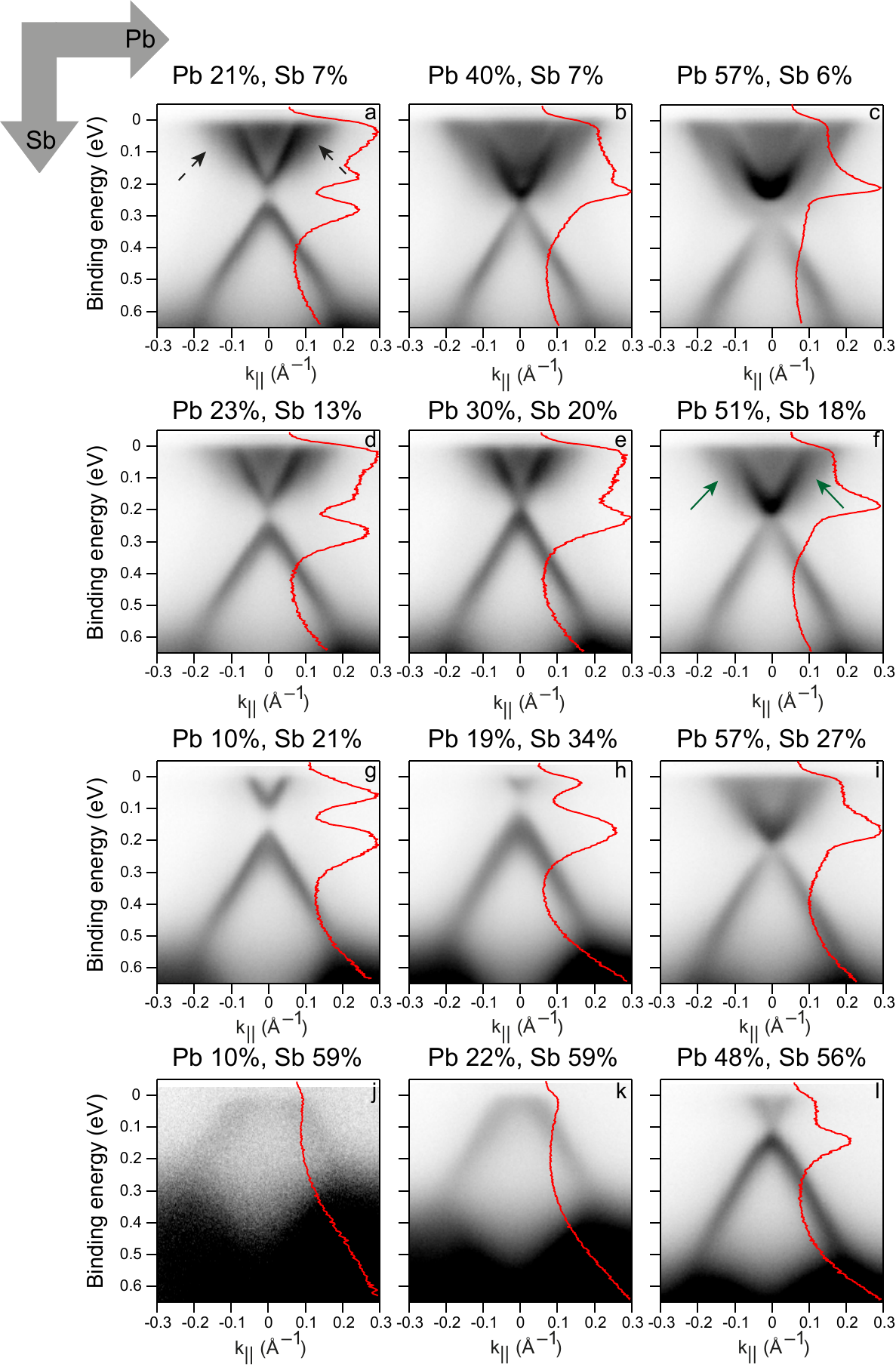}
\caption{\label{fig1} The dispersion relations for \Pyat~samples with varying Pb-values and Sb-values ($h\nu=21.2$~eV). The molar fractions of Pb and Sb are shown at the top of the panels. The Pb concentration increases from left to right, and the Sb concentration increases from top to bottom.}
\end{figure*}

To systematically study the effect of composition on the electronic structure, a series of approximately 40 samples of the \Pyat~system with varying Pb and Sb concentrations was investigated. The elemental composition of each sample was determined using energy-dispersive X-ray spectroscopy and X-ray photoelectron spectroscopy. The position of the Dirac point, designated as $E_D$, relative to the Fermi level, and the value of the bulk band gap, denoted here as $E_g$, were measured using ARPES. The full data set is provided in the Supplementary Material.

Fig.~\ref{fig1} shows the electronic band dispersion of the \Pyat~system for different Pb and Sb concentrations. The spectra are arranged with increasing Pb concentration along the horizontal axis and Sb concentration along the vertical axis. On the right side of each panel, the energy distribution curves measured at the $\Gamma$-point ($k=0$) are shown in red.

Substituting Bi with Sb leads to p-type doping, which is characterized by a shift of the electronic structure toward lower binding energies. The position of the Dirac point $E_D$ relative to the Fermi level changes, but its dependence on the Sb concentration is not strictly monotonic, necessitating a separate analysis (see Table~\ref{tab1}). Increasing the Pb concentration at a constant Sb concentration decreases the bulk band gap value until it closes at Pb concentrations of approximately $40-50$\%, as illustrated in Fig.~\ref{fig1}~b. At the same time, Rashba-like states emerge and intensify (indicated by arrows in Fig.~\ref{fig1}~a). As the Pb concentration increases, these states overlap the upper part of the Dirac cone and shift toward higher binding energies (see Fig.~\ref{fig1}~c). The effect of Pb on the enhancement of Rashba-like states is most noticeable at low Sb concentrations ($\sim7\%$). As Sb increases ($>20\%$), the Rashba-like states become less intense, and their intensity ceases to increase with further Pb addition. The observed differences correlate with the decreasing Bi content, suggesting that Rashba-like states are primarily associated with Bi atoms. However, their origin in this system may also be related to band bending, and the exact mechanism remains an open question.

The values of the bulk band gap $E_g$ and the position of the Dirac point $E_D$ for each sample are listed in Table~\ref{tab1}. The Dirac point positions above the Fermi level are obtained by extrapolating the branches of the Dirac cone from the corresponding ARPES spectra. At low Sb concentrations ($\approx7\%$, top row of Fig.~\ref{fig1}~a–c), the transition from Pb~$20\%$ to Pb~$40\%$ is accompanied by band gap closure – $E_g$ decreases from 86 to 0~meV. At intermediate Sb concentrations ($13$--$34\%$, rows d–i), the bulk band gap decreases to 44~meV for the sample with Pb~$30\%$ and Sb~$20\%$ (Fig.~\ref{fig1}~e). For the sample with Pb~$51\%$ and Sb~$18\%$ (Fig.~\ref{fig1}~f), the band gap closes, and the states above the Dirac point (shown by green arrows) broaden along the wave vector. The electronic structure here differs from that of the sample with Sb~$6\%$ (Fig.~\ref{fig1}~c), where the broadening was more pronounced.

\begin{table}
\caption{\label{tab1}%
Concentrations of Pb and Sb and the corresponding values of the bulk gap value (E$_g$) and the position of the Dirac point (E$_D$), defined as $E-E_F$, relative to the Fermi level.}
\begin{ruledtabular}
\begin{tabular}{cccc}
\textrm{Pb (\%)}&
\textrm{Sb (\%)}&
\multicolumn{1}{c}{E$_{G}$~(meV)}&
\textrm{E$_{D}$~(meV)}\\
\colrule
10 & 59 & -- & -100\\
22 & 59 & -- & -60\\
48 & 56 & 0 & 80\\
19 & 34 & 51 & 84\\
10 & 21 & 109 & 138\\
30 & 20 & 43 & 180\\
57 & 27 & 0 & 214\\
23 & 13 & 83 & 216\\
51 & 18 & 0 & 216\\
21 & 7 & 86 & 232\\
40 & 7 & 0 & 266\\
57 & 6 & 0 & 300\\
\end{tabular}
\end{ruledtabular}
\end{table}

The position of the Dirac point $E_D$ in the \Pyat~system depends on the concentrations of Pb and Sb. The effect of Pb can be assessed by examining two samples with similar Sb concentrations ($\approx6-7\%$) but significantly different Pb contents. In the sample with Pb $21\%$ and Sb~$7\%$ (Fig.~\ref{fig1}~a) the Dirac point is located at a binding energy of 232~meV, whereas for Pb~$57\%$ and Sb~$6\%$ (Fig.~\ref{fig1}~c) $E_D=300$~meV. An almost twofold increase in the Pb content leads to a noticeable shift of the Dirac point, indicating the significant role of Pb in determining the position of the Fermi level. At high Sb concentrations ($\sim60\%$, bottom row, j–l), the system remains p-type up to Pb~$48\%$, at which point a transition to n-type conductivity occurs. Consequently, the position of the Fermi level is influenced not only by Sb but also by Pb concentration.

Machine learning methods were used to establish the functional dependence of $E_D$ and $E_g$ on composition. The Random Forest algorithm, implemented in Python, was trained on data from all samples. The input features were the molar fractions of Pb ($x$) and Sb ($y$), ranging from 0 to 1, as well as their nonlinear combinations: $x^2$, $y^2$, and the ratio $\frac{x}{y}$ (or $\frac{y}{x}$). Analysis of the feature importance matrix from the trained model allowed a quantitative assessment of the contribution of each factor. It was found that the value of the bulk band gap $E_g$ is almost entirely determined by the Pb concentration: the relative importance of the Pb concentration was $\approx$0.75, while the contribution of the other features (including the linear concentration of Sb) was much smaller.

A different dependence is found for the Dirac point position $E_D$. Unexpectedly, the dominant factor turned out to be not the Sb concentration but the concentration ratio $\frac{x}{y}$ (or equivalently $\frac{y}{x}$) — its importance was approximately 0.7–0.85. Constructing a model based exclusively on the linear parameters of Pb and Sb without introducing nonlinear combinations reveals the expected dominant role of Sb concentration. However, the inclusion of cross-terms shows that the Fermi level position in this system is governed by the balance of Pb and Sb cations, rather than by the absolute Sb content.

\begin{figure}
\includegraphics[width=0.45\textwidth]{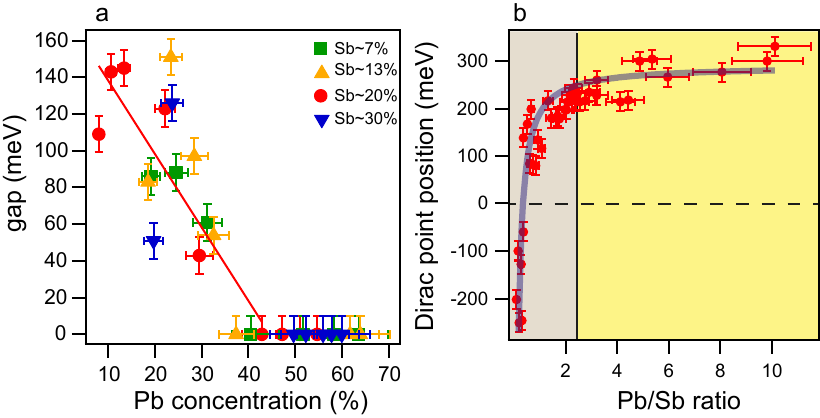}
\caption{\label{fig2} (a) Dependence of the bulk band gap value on Pb concentration. (b) Dependence of the Dirac point position on the Pb/Sb ratio. The blue solid line represents the fit.}
\end{figure}

Fig.~\ref{fig2}~a shows the dependence of the band gap $E_g$ on the Pb concentration (data from Table~\ref{tab1}). The experimentally obtained values follow a linear dependence on the Pb concentration with little scatter. Different symbols correspond to different Sb concentrations. Regardless of the Sb concentration, an increase in Pb leads to a decrease in $E_g$, and at Pb concentrations of $\approx 40-50\%$ the gap closes. In the range of Pb concentrations from 0 to 50\%, the dependence is nearly linear; with a further increase in Pb to 60\%, $E_g$ remains close to zero. This behavior indicates a transition of the system to a topological phase distinct from the TI phase. It is consistent with the results of studies on the substitution of Mn with Pb in \MBT~\cite{qian2022magnetic,estyunin2025electronic}. The absence of a noticeable effect of Sb concentration on the band gap value indicates that its closure is determined not so much by spin-orbit interaction as by electrostatic effects and the redistribution of electron density at the band edges. Since spin-orbit interaction strongly depends on the atomic number, the substitution of Bi with Sb would have significantly altered the band gap width if this mechanism were dominant. The fact that $E_g$ does not depend on the Sb concentration implies that the decisive role is played by changes in ionic radii, interatomic distances, and Coulomb interactions upon the substitution of Mn with Pb \cite{qian2022magnetic}, as well as the associated redistribution of electron density at the edges of the band gap.

Fig.~\ref{fig2}~b shows the dependence of the Dirac point position on the Pb/Sb ratio. All experimental data for $E_D$ are well described by a function of the Pb/Sb ratio. As this ratio increases, the Dirac point shifts toward higher binding energies. In contrast to the substitution of Bi with Sb in \MBT, where the position of the Dirac point is determined solely by the Sb concentration, in the doubly substituted system both elements contribute to the Dirac point position.

Two distinct intervals can be identified in the data, where the Pb/Sb ratio behaves differently (shown in grey and yellow in Fig.~\ref{fig2}~b). At low values of the ratio, corresponding to the region where the Sb concentration predominates, the dependence is quite sharp: a small increase in the ratio leads to a significant shift of the Dirac point position. This indicates a high sensitivity of the electronic structure to changes in the Pb/Sb ratio in this region. At high Pb/Sb values, corresponding to the predominance of Pb concentration, the curve becomes flatter, and within the studied range, a further increase in Pb concentration has almost no effect on the position of the Dirac point.

The dependence of the Dirac point position on the Pb concentration is explained as follows. As shown in the work by Qian et al. \cite{qian2022magnetic} for the (Mn,Pb)Bi$_2$Te$_4$ system, the charge carrier concentration increases by nearly an order of magnitude with increasing Pb concentration (see Table I in the cited article). Consequently, at high Pb concentrations, increasing amounts of Sb are required to compensate for the electron doping, which is achieved through the formation of cation vacancies or other defects that effectively reduce the electron concentration. This also helps position the Fermi level inside the band gap; therefore, the sensitivity of $E_D$ to the Pb/Sb ratio decreases, and the curve reaches a plateau.

The transition from n-type to p-type conductivity occurs at a Pb/Sb ratio of $\sim$0.5. This value corresponds to the compensation point between electron (related to Pb substitution) and hole (related to Sb) doping, and the Fermi level lies near the Dirac point. The fits were obtained for ratios above 0.05 (i.e., for compositions containing both Pb and Sb). For very small Pb/Sb ratios (close to zero, corresponding to nearly pure \MBT), deviations from the general dependence may occur. However, for Pb/Sb > 0.05, the function agrees well with the experimental data (see Fig.~\ref{fig2}~b), confirming that the Dirac point position in the studied concentration range is determined by the Pb/Sb ratio.

\subsection{$\mu$-ARPES data ($h\nu=6.3$~eV)}

\begin{figure*}
\includegraphics[width=0.8\textwidth]{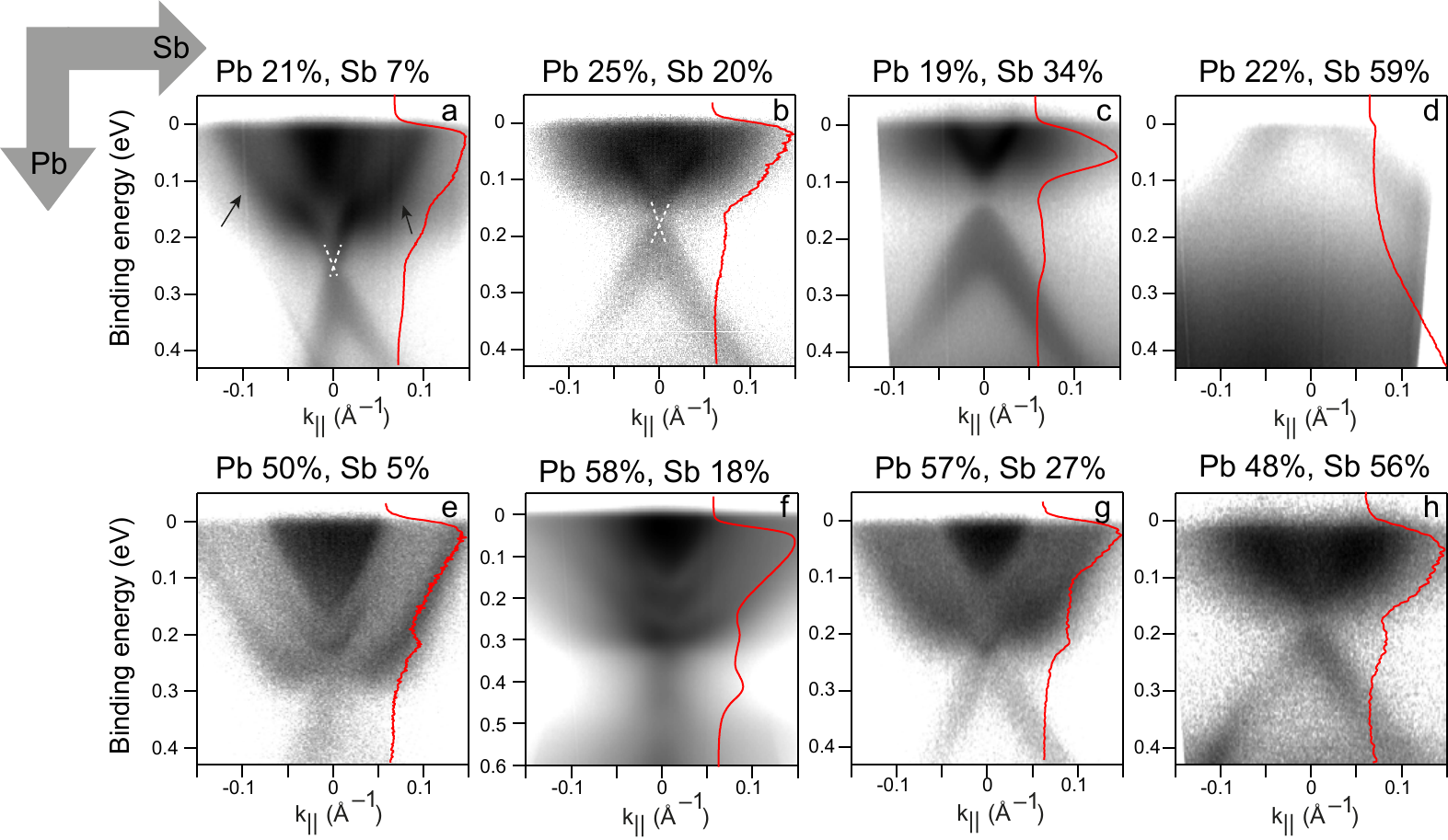}
\caption{\label{fig3} The dispersion relations for \Pyat~samples with varying Pb and Sb concentration ($h\nu=6.3$~eV). The molar fractions of Pb and Sb are shown at the top of the panels.}
\end{figure*}

The presence of TSS in the electronic structure is a criterion for the topological phase of material. For \MBT~at a photon energy of 21.2~eV, the photoemission spectra are mainly determined by bulk-related contributions, while the surface states are weakly contributing, making it difficult to identify TSS. Therefore, additional measurements were performed at a photon energy of $h\nu = 6.3$~eV for the samples shown in Fig.~\ref{fig1}. The experimental data are presented in Fig.~\ref{fig3}. On the right of each panel, the energy distribution curves (shown in red) taken at the $\Gamma$-point are plotted.

In the region of low Sb concentrations (Fig.~\ref{fig3}~a,e), at 21\% Pb and 7\% Sb (Fig.~\ref{fig3}~a), the TSS are well resolved (shown by the dashed line), and Rashba-like states (black arrows) have higher intensity than those measured at 21.2~eV. As the Pb content increases to 50\% (Sb 5\%, Fig.~\ref{fig3}~e), the Rashba states shift toward higher binding energies; at the same time, the TSS broaden, making their identification more difficult. These data are consistent with a topological phase transition at Pb concentrations of about 40–50\% \cite{estyunin2025electronic}. The system enters a phase distinct from the TI state, and the bulk band gap closes.

At Pb concentrations of 25\% and Sb~20\% (Fig.~\ref{fig3}~b), TSS are observed, but the Rashba-like states are less pronounced than at lower Pb concentrations. At a Pb concentration of 58\% and Sb 18\% (Fig.\ref{fig3}~f), it becomes difficult to distinguish the TSS, which corresponds to a topological phase transition at Pb$\sim$40–50\%. Consistent with earlier studies \cite{ma2021realization}, the increase in Sb concentration at low Pb opens a gap in the TSS.


At Pb concentrations of 19\% and Sb concentrations of 34\% (Fig.~\ref{fig3}c), the electronic structure resembles the data obtained for \MBT~when Bi is substituted with Sb~\cite{ma2021realization}. Thus, at this concentration, Pb does not significantly affect the electronic structure; its role is mainly to shift the Rashba-like states to higher binding energies. At a Pb concentration of 57\% and an Sb concentration of 27\% (Fig.~\ref{fig3}~g), the TSS cannot be unambiguously identified in the spectrum.

At Pb~22\% and Sb~59\% (Fig.~\ref{fig3}~d), the electronic structure exhibits p-type conduction, which makes it difficult to identify the TSS. For the sample with Pb~48\% and Sb~56\% (Fig.~\ref{fig3}~h), a strong broadening of the spectral lines is observed. This spectrum resembles the data for Pb~57\% and Sb~27\% (Fig.~\ref{fig3}~g), except that the electronic structure is shifted to lower binding energies. In both cases, clear identification of the TSS in the presence of bulk states is difficult. Therefore, additional techniques are required for unambiguous determination of the topological phase, such as circular dichroism ARPES (CD-ARPES), which allows the TSS to be identified by the characteristic CD signal.

\subsection{Circular dichroism}

In CD-ARPES, the difference between spectra obtained with right- and left- circular polarization light is analyzed. Correlated with the helical spin structure of TSS, in which the electron spin is perpendicular to the wave vector, the branches of the Dirac cone with opposite spin directions yield opposite signs of the CD signal relative to the $\Gamma$-point (in the figures, conventionally shown in red and blue) \cite{wang2013circular}. Thus, CD-ARPES allows one to distinguish TSS from bulk states; however, it does not provide direct information about the spin structure.

\begin{figure*}
\includegraphics[width=0.75\textwidth]{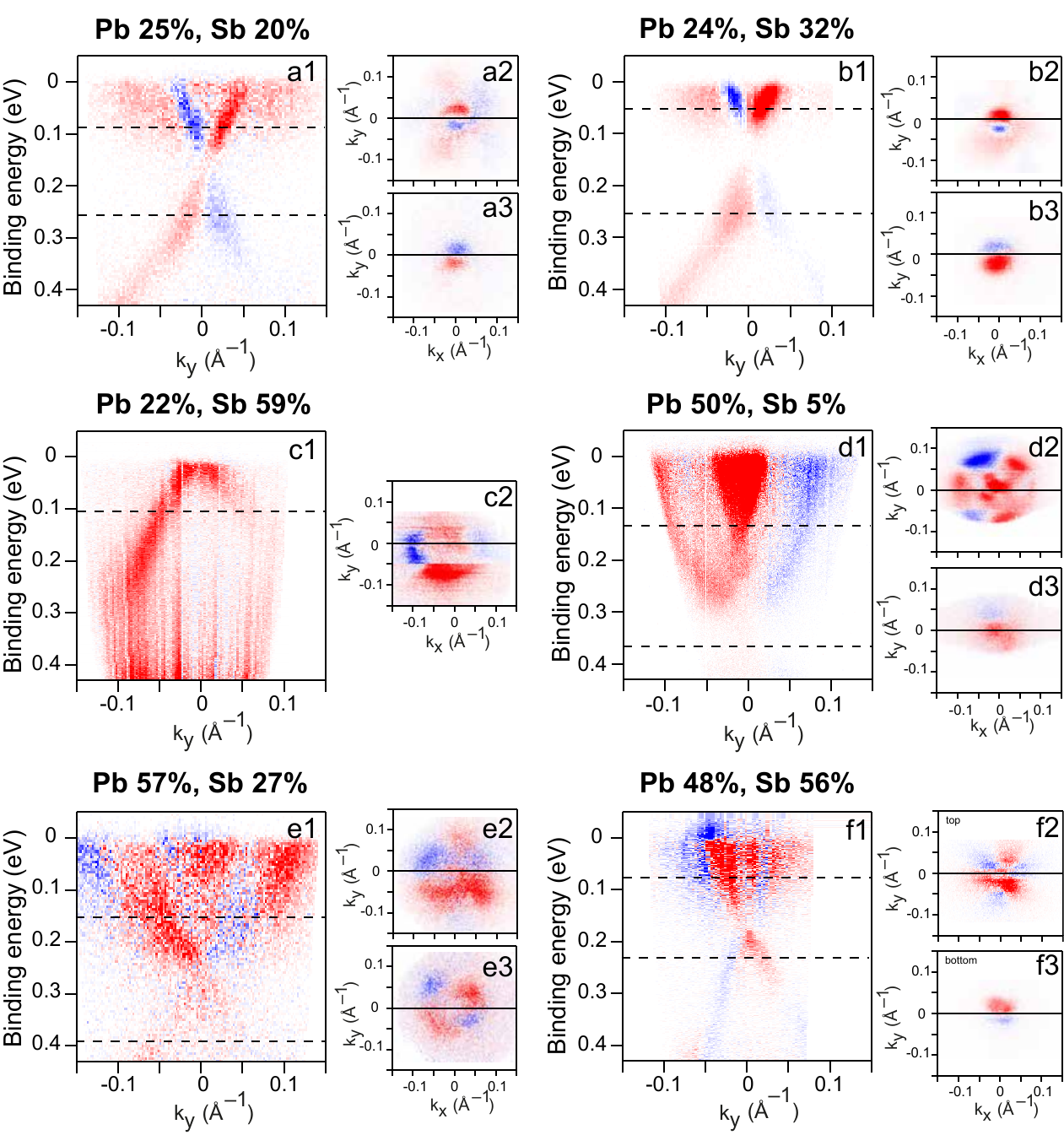}
\caption{\label{dich} Circular dichroism ARPES spectra of \Pyat~at various Pb and Sb concentrations.}
\end{figure*}

Figure~\ref{dich} presents data obtained using the CD-ARPES method for the same samples as in Fig.~\ref{fig3}. On the left side of each panel is the I(E,k$_y$) spectrum in a direction perpendicular to the analyzer slit. The right side shows the I(k$_x$,k$_y$) curves for constant energy cuts in the conduction band region (shown at top) and the valence band region (bottom).


For samples with a Pb concentration of 25\% and an Sb concentration of 20\% (Fig.~\ref{dich}~a), the TSS are clearly distinguishable in both dispersion and constant-energy cuts - a sign inversion of the difference intensity (blue $\rightarrow$ red) is observed when crossing the plane of incident light. For the sample with 32\% Sb and 24\% Pb (Fig.~\ref{dich}~b), the TSS are clearly visible. For this sample, as well as for others with similar low Pb and moderate Sb, the TSS are well resolved both above and below the Dirac point, maintaining the characteristic CD sign reversal relative to $\Gamma$-point. This CD signal persists upon Bi‑for‑Sb substitution, indicating that the TSS remain clearly discernible. Similar behavior was observed in Sb-doped \MBT~by Ma et al.~\cite{ma2021realization}, and here we find that it also holds in the doubly substituted system for samples with low Pb content (e.g., Fig.~\ref{dich}~a, b).

In samples with 50\% Pb, 5\% Sb (Fig.~\ref{dich}~d) and 57\% Pb, 27\% Sb (Fig.~\ref{dich}~e), the characteristic CD signal of TSS is absent. The lower part of the cone in the I(E,k) spectra is weak or masked by the background; in the valence band cuts, no sign inversion is observed at $k=0$. This indicates that at Pb concentrations of about 40\% and higher, the system transitions to a phase different from the TI. The discrepancy between the observed threshold (Pb~40\% instead of the expected 50\% \cite{estyunin2025electronic}) may be due to errors in concentration determination.

For samples with a high Sb concentration (approximately 60\%, panels c and f), strong p-doping complicates the interpretation. In the sample with 22\% Pb (Fig.~\ref{dich}~c), only valence band states (below the Dirac point) are observed. In the sample with 48\% Pb (Fig.~\ref{dich}~f), both branches of the cone are visible with distinct CD signal.

Thus, for all studied Sb concentrations (5–60\%), the system remains in the TI phase when the Pb content does not exceed 40\%. The CD signal clearly indicates that the states exhibiting this signal are TSS. Raising Pb above 40\% triggers a topological phase transition, which depends weakly on the Sb concentration. Substituting Bi with Sb modulates the Dirac point position relative to the Fermi level and affects the spectral details; however, the CD signal that allows identification of TSS is preserved. According to detailed studies \cite{ma2021realization}, the spin polarization of TSS in such systems may vary, but the states themselves remain topologically protected and recognizable by CD contrast.

\subsection{Magnetic structure}

\begin{figure*}
\includegraphics[width=0.8\textwidth]{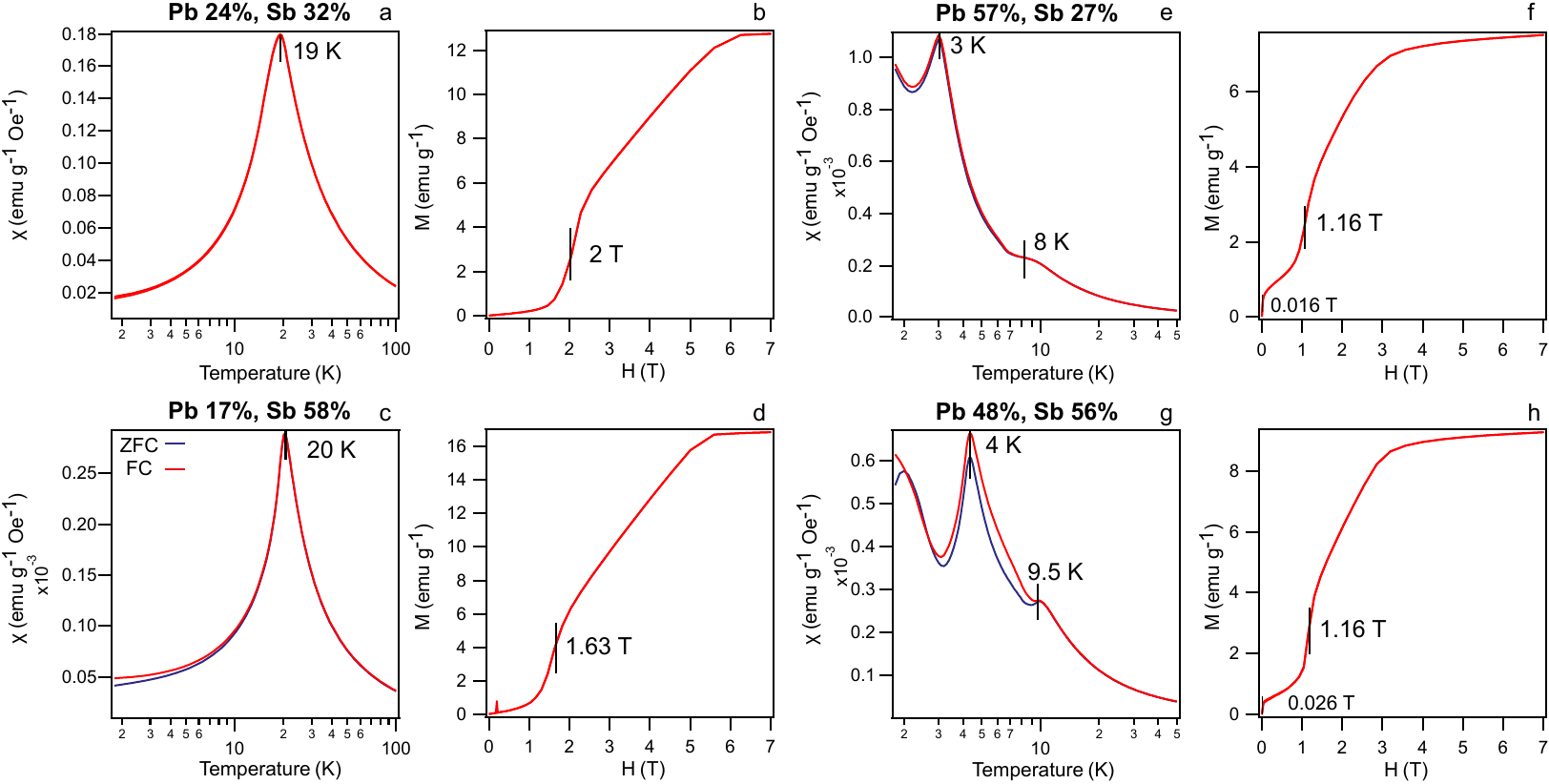}
\caption{\label{mag} Temperature dependencies of the magnetic susceptibility $\chi(T)$ and magnetization as a function of applied magnetic field, M(H) were measured for \Pyat~samples using a SQUID magnetometer. The $\chi(T)$ measurements were performed with an external field of 5~mT applied perpendicular to the sample surface along the crystallographic $c$ direction. The blue curve in panels a, c, e, g corresponds to zero-field-cooled (ZFC) and the red curve to field-cooled (FC) conditions.}
\end{figure*}

The magnetic properties of (Mn,Pb)Bi$_2$Te$_4$ and Mn(Bi,Sb)$_2$Te$_4$ have been studied in detail separately (see, for example, \cite{qian2022magnetic,estyunin2023comparative,yan2019evolution}). To investigate the magnetic order in the \Pyat~system, we performed magnetic measurements. The results for the studied samples are shown in Fig.~\ref{mag}.

To analyze the effect of substitution on the magnetic properties, the samples were grouped into two pairs based on their Pb content (left and right halves of Fig.~\ref{mag}). Within each pair, the samples have similar Pb concentrations but differ significantly in Sb content.

The left half of Fig.~\ref{mag} presents the curves for the sample with 24\% Pb and 32\% Sb (top row, Fig.~\ref{mag}a,b) and for the sample with 17\% Pb and 58\% Sb (bottom row, Fig.~\ref{mag}c,d). The temperature dependencies of $\chi(T)$ for both samples show a sharp peak corresponding to the Néel temperature $T_N = 19$~К for the first sample and $T_N = 20$~К for the second. This behavior is characteristic of antiferromagnetic ordering. The absence of significant discrepancies between the curves measured in FC and ZFC regimes indicates that the samples do not contain crystalline phases other than the \MBT‑type phase (considering Mn/Pb and Bi/Sb substitutions). The obtained $T_N$ values are consistent with previously reported data for Mn$_{1-x}$Pb$_x$Bi$_2$Te$_4$ samples with Pb$\approx$20\% \cite{estyunin2023comparative}. Notably, the Néel temperatures of the two samples are close (19~K and 20~K), despite a large difference in Sb concentration (32\% vs 58\%) and the fact that the first sample is n‑type, while the second is p‑type. This indicates that variations in Sb concentration within the studied range have little effect on the magnetic properties, even though they significantly alter the electronic structure, specifically the conductivity type. Figures~\ref{mag}~b and d show the magnetization as a function of magnetic field $M(H)$. The behavior of these curves is similar for both samples. The spin‑flop transition field is 2~T for the 24\% Pb, 32\% Sb sample and 1.63~T for the 17\% Pb, 58\% Sb sample. The spin‑flop fields differ slightly more than the $T_N$ values, but the difference remains small.

For the second pair of samples (the right side of Fig.~\ref{mag}), the concentrations are Pb 57\%, Sb 27\% (Fig.~\ref{mag}~e, f) and Pb 48\%, Sb 56\% (Fig.~\ref{mag}~g, h). The temperature dependencies of the magnetic susceptibility $\chi(T)$ for these samples differ from those discussed above: two peaks are observed. The sample with Pb 57\%, Sb 27\% (Fig.~\ref{mag}~e) has a Néel temperature of 8~K, which agrees well with reported values for Mn$_{1-x}$Pb$_x$Bi$_2$Te$_4$ at similar Pb concentrations. For the sample with Pb 48\%, Sb 56\% (Fig.~\ref{mag}~g), $T_N = 9.5$~К. In addition to the peaks associated with the \MBT~phase, both curves show additional features at 3~K and 4~K. These additional peaks may arise from sample imperfections and impurity phases other than the main \Pyat~compound. Apparently, they are caused by the impurity phase MnBi$_4$Te$_7$, for which the magnetic ordering temperature is approximately half that of \MBT~(considering Mn/Pb and Bi/Sb substitutions) \cite{klimovskikh2020tunable}. Within experimental error, the temperatures of the additional transitions (3~K and 4~K) are close, and no significant dependence on the Sb concentration is observed. The M(H) curves for both samples (Fig.~\ref{mag}~f,~h) show a spin‑flop transition at 1.16~T; the transition fields coincide within experimental error for both samples.



The results show that substituting Bi for Sb, which alters the electron density (and thus the electronic structure), has virtually no effect on the magnetic properties. This suggests that exchange interactions depend weakly on the electronic structure near the Fermi level, which does not rule out localized mechanisms (e.g., superexchange). However, further studies are required to draw a reliable conclusion. In contrast, the substitution of Mn with Pb leads to a dilution of the magnetic sublattice, causing a monotonic decrease in $T_N$ and $H_{sf}$. Consequently, the concentration of magnetic Mn ions plays a key role in the magnetism of the studied systems.

\section{Conclusion}

In this work, (Mn,Pb)(Bi,Sb)$_2$Te$_4$ crystals with Pb concentrations of 10–60\% and Sb concentrations of 5–60\% were studied using ARPES, circular dichroism, and SQUID magnetometry. The results show that the position of the Dirac point is determined not by the absolute Sb concentration but by the Pb/Sb ratio: at low values of this ratio (Sb dominance), the Dirac point shifts strongly to higher binding energies, while at high values (Pb dominance) it tends to saturate. The transition from p‑type to n‑type conduction occurs at Pb/Sb $\approx$ 0.5, which corresponds to a compensated semiconductor.

The topological phase transition, accompanied by the closure of the bulk band gap and the disappearance of the TSS, occurs at Pb $\sim$40–50\% irrespective of the Sb concentration. Substituting Mn with Pb induces this transition, whereas substituting Bi with Sb does not affect the topological phase but allows tuning of the Dirac point position. TSS are preserved for Pb < 40\%, while at Pb $\sim$50\%, states with the characteristic TSS signature cannot be resolved, confirming the transition from the TI phase.

Magnetic measurements revealed that antiferromagnetic order is preserved in \Pyat~across  the entire studied composition range (x = 10–60\%, y = 5–60\%). For samples containing $\sim$20\% Pb (regardless of the Sb content: 32\% or 58\%), the Néel temperature is 19–20~K, and the spin‑flop transition field is 1.6–2~T. For samples with Pb~50\% (Sb 27\% and 56\%), the Néel temperature is 8–9.5~K (with an additional feature at 3–4~K, likely associated with an secondary phase), and the spin‑flop transition field is 1.16~T. In each pair of samples with similar Pb concentrations but significantly different Sb contents, the magnetic characteristics are practically identical. This implies that the substitution of Bi for Sb does not affect the magnetic sublattice, and the exchange mechanism is consistent with Mn–Te–Mn superexchange, which is insensitive to changes in the electronic environment at the Bi/Sb sites even with significant dilution of the magnetic sublattice by Pb atoms.

Thus, the simultaneous substitution of Mn with Pb and Bi with Sb in \MBT~allows independent control of the topological phase (via the Pb concentration) and the position of the Dirac point (via the Pb/Sb ratio) while preserving antiferromagnetic order across the entire composition range studied. By varying the Pb/Sb ratio, the Dirac point can be tuned to the Fermi level without changing the topological phase, which opens the possibility of independently tuning electronic and magnetic properties in search of materials with non‑trivial topology and controllable electronic structure features.

\section{Experimental section}

\textbf{Sample Synthesis:}

Crystal growth was carried out using a modified Bridgman method. A Bi$_2$Te$_3$-based flux was employed, to which a stoichiometric charge of the target compound was added. All components were used in elemental form with high purity. The materials were loaded into a quartz ampoule with a conical bottom, internally coated with pyrolytic carbon. After evacuation to a vacuum level of $5 \times 10^{-3}$~Torr, the ampoule was sealed using a propane–oxygen torch. Recrystallization was performed in a stationary temperature gradient of a two-zone furnace while translating the ampoule at a rate of 9~mm/day.

\textbf{ARPES measurements:}

ARPES data were recorded using (i) laboratory-based facility SPECS GmbH ProvenX-ARPES system located in ISP SB RAS (Novosibirsk, Russia) equipped with ASTRAIOS 190 electron energy analyzer, a non-monochromated He I~$\alpha$ light source with h$\nu$=21.22~eV and sample cooling by liquid nitrogen (Fig.\ref{fig1}); (ii) $\mu$-Laser-ARPES facility at Research Institute for Synchrotron Radiation Science (HiSOR) at Hiroshima University (Japan) equipped with Scienta R4000 analyzer, laser light source with h$\nu$=6.3~eV and sample cooling by liquid helium (Fig.\ref{fig3},~\ref{dich}) \cite{iwasawa2017development}. A clean surface was prepared by tape-cleavage inside a vacuum chamber with a pressure below $10^{-8}$ torr. The base pressure during the measurements remained below $5\times10^{-11}$ torr.

\textbf{SQUID magnetometry:}

Magnetic properties (i.e. the temperature dependences of the magnetic susceptibility $\chi$(T) and magnetization as a function of applied magnetic field M(H) at various temperatures) were assessed using a Quantum Design MPMS SQUID VSM instrument. The measurements were conducted at temperatures as low as 1.8~K, with an external magnetic field range of $\pm7$T and a sensitivity of $1\times10^{-8}$~emu at 0~T. The samples were oriented such that the external magnetic field was applied perpendicular to the crystal surface (0001), aligning with the crystallographic direction c.

\begin{acknowledgments}

This work was supported by the St. Petersburg State University grant No. 125022702939-2.

ARPES measurements at HiSOR were performed under Proposal No. 25AG012. We are grateful to the N-BARD, Hiroshima University for liquid He supplies. The authors acknowledge the following centers of the Research Park of St. Petersburg University: the ``Centre for Nanotechnology'', where the elemental composition of the samples was studied and the “Centre for Diagnostics of Functional Materials for Medicine, Pharmacology and Nanoelectronics”, where the magnetic properties of the materials were studied. Crystal growth was performed under state assignment of IGM SB RAS FWZN-2026-0005. O.E.T. and V.A.G. acknowledge support from the SRF SKIF Boreskov Institute of Catalysis (No. FWUR-2024-0040) and from ISP SB RAS.

The manuscript was written and edited by T.P.M., D.A.E. and A.M.S. Experimental data processing and figure preparation by T.P.M. and D.A.E. ARPES measurements at ISP SB RAS were performed by T.P.M., D.A.E., A.S.F., V.A.G., and O.E.T. ARPES measurements at HiSOR facilities ($\mu$-ARPES) were performed by T.P.M., D.A.E., A.S.F., Y.K. and K.S. Crystals provided by K.A.K. and O.E.T. Funding was provided by A.M.S. The project was planned and supervised by T.P.M., D.A.E. and A.M.S.

\end{acknowledgments}

\appendix



\providecommand{\noopsort}[1]{}\providecommand{\singleletter}[1]{#1}%

\end{document}